\documentclass{jfm}

\usepackage{graphicx}
\usepackage{natbib}
\usepackage{color}
\usepackage{xcolor}

\ifCUPmtlplainloaded \else
  \checkfont{eurm10}
  \iffontfound
    \IfFileExists{upmath.sty}
      {\typeout{^^JFound AMS Euler Roman fonts on the system,
                   using the 'upmath' package.^^J}%
       \usepackage{upmath}}
      {\typeout{^^JFound AMS Euler Roman fonts on the system, but you
                   dont seem to have the}%
       \typeout{'upmath' package installed. JFM.cls can take advantage
                 of these fonts,^^Jif you use 'upmath' package.^^J}%
      }
  \else
  \fi
\fi

\ifCUPmtlplainloaded \else
  \checkfont{msam10}
  \iffontfound
    \IfFileExists{amssymb.sty}
      {\typeout{^^JFound AMS Symbol fonts on the system, using the
                'amssymb' package.^^J}%
       \usepackage{amssymb}%

      }{}
  \fi
\fi

\ifCUPmtlplainloaded \else
  \IfFileExists{amsbsy.sty}
    {\typeout{^^JFound the 'amsbsy' package on the system, using it.^^J}%
     \usepackage{amsbsy}}
    {\providecommand\boldsymbol[1]{\mbox{\boldmath $##1$}}}
\fi

\newsavebox{\astrutbox}
\sbox{\astrutbox}{\rule[-5pt]{0pt}{20pt}}

\title[A focused liquid jet formed by a water hammer in a test tube]{A focused liquid jet formed by a water hammer in a test tube}

\author[A. Kiyama, Y. Tagawa, K. Ando and M. Kameda]%
{Akihito Kiyama$^1$,\ns
Yoshiyuki Tagawa$^1$\thanks{Email address for correspondence: tagawayo@cc.tuat.ac.jp}, \ns
Keita Ando$^2$ \ns
and \\Masaharu Kameda$^1$\break}

\affiliation{$^1$Department of Mechanical Systems Engineering, Tokyo University of Agriculture and Technology,
Nakacho 2-24-16 Koganei, Tokyo 184-8588, Japan\\
$^2$Department of Mechanical Engineering, Keio University, 
3-14-1 Hiyoshi, Kohoku-ku, Yokohama, Kanagawa 223-8522, Japan\\[\affilskip]}

\pubyear{2010}
\volume{650}
\pagerange{119--126}
\date{?; revised ?; accepted ?. - To be entered by editorial office}
\begin{document}

\maketitle

\begin{abstract}
We investigate motion of a gas-liquid interface in a test tube induced by a large acceleration via impulsive force.
We conduct simple experiments in which the tube partially filled with a liquid falls under gravity and impacts a rigid floor. 
A curved gas-liquid interface inside the tube reverses and eventually forms an elongated jet (i.e. the so-called a focused jet).
In our experiments, there arises either vibration of the interface or increment in the velocity of a liquid jet accompanied by the onset of cavitation in the liquid column.
These phenomena cannot be explained by considering pressure impulse in a classical potential flow analysis, which does not account for finite speeds of sound as well as phase change.
Here we model such water-hammer events as a result of one-dimensional pressure wave propagation and its interaction with boundaries through acoustic impedance mismatching.
The method of characteristics is applied to describe pressure wave interactions and the subsequent cavitation.
The proposed model is found to allow us to capture the unsteady features of the liquid jet.
\end{abstract}

\section{Introduction}
\label{I}
A liquid jet is of great importance in various industrial and medical processes as well as of fundamental interest as a canonical fluid dynamical phenomenon to study the instability in motion of gas-liquid interfaces \citep{Eggers2008,Duchemin2008,Bartolo2006,Bergmann2008,Tagawa2013}.

\begin{figure}
\centerline{\includegraphics[width=1.0\columnwidth]{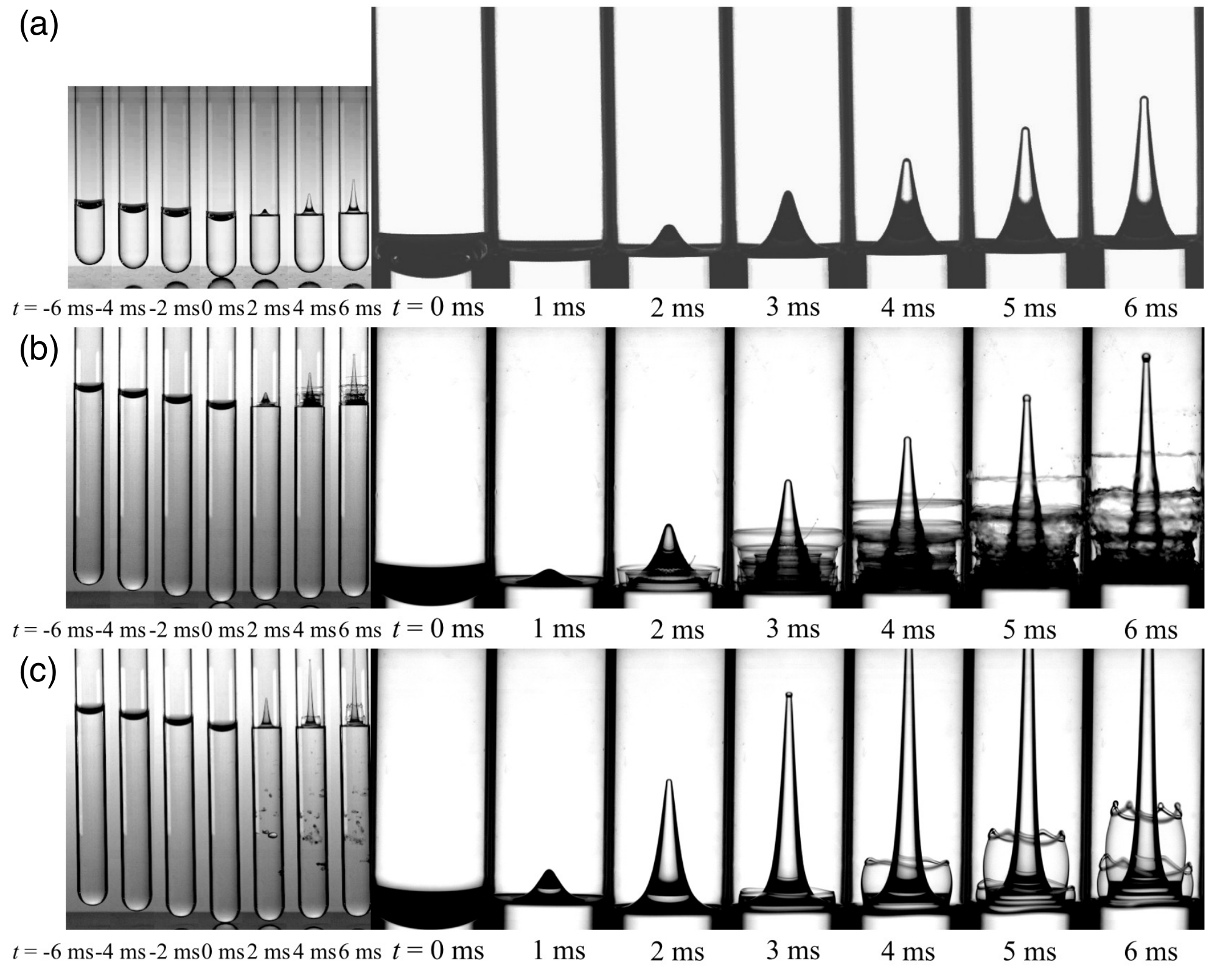}}
  \caption{Three typical types of jets: (a) Normal type (also see supplementary movie 1). This jet can be described by the (incompressible) pressure impulse. (b) Splash type (supplementary movie 2). Non-trivial vibration of the interface with small droplets sprayed is observed. (c) Cavitation type (supplementary movie 3). Its jet velocity is much faster than that of the other 2 types.}
\label{type}
\end{figure}

One of the typical jets is a jet whose tip is sharp and elongated, i.e., the so-called  ``focused jet" \citep{Eggers2008, Tagawa2012, Peters2013}.
Such a focused jet can be created as follows:
A container partially filled with a liquid is instantly set into motion, thus letting all the fluid particles that include the gas-liquid interface be under rapid acceleration.
The accelerated interface then deforms into a focused jet by flow focusing effect, termed as ``shaped-charge effect''\citep{Birkhoff1948}.
One of the representative examples is Pokrovski's experiment \citep{Antkowiak2007}(See Figure \ref{type}a, Supplementary movie 1).
In this experiment, a liquid-filled test tube falls freely and eventually collides with a rigid floor.
During the tube's free-fall (i.e., in the gravity-free state), a gas-liquid interface quickly deforms as hemispherical shape by its surface tension. 
Once the tube impacts the floor, the direction of its motion reverses.
The acceleration of fluids within the tube leads to the formation of a focused jet from the interface.

It has been known that a flow triggered by sudden motion of boundaries can be analyzed by considering \textit{pressure impulse} \citep{Batchelor1967,Cooker2006} defined as the time integral of pressure evolution.
The potential theory assumes an instantaneous establishment of pressure fields through the infinite speed of sound, meaning that the characteristic length of acoustic waves is assumed much larger than fluid-dynamic length scales to validate the incompressibility condition.
Using this pressure impulse approach, \citet{Antkowiak2007} analyzed the velocity field right after the impact in Pokrovski's experiment and obtained a good agreement with their experiments.
\citet{Kiyama2014} conducted similar experiments and found that the jet velocity can be described by the pressure impulse approach as well.   
The jet velocity can be written as $V_j = \alpha U_{0}$, where $U_0$ is the velocity of the gas-liquid interface just after the impact and $\alpha$ is a dimensionless constant to be determined empirically.
The physical meaning of $\alpha$ is the strength of flow focusing effect after the interface obtains the value of velocity $U_0$.

However, 
as the impact is enlarged, we find that motion of a gas-liquid interface tends to deviate from the previous findings:
Non-trivial vibration of the interface with droplet fragments sprayed (See Figure \ref{type} (b), Supplementary movie 2), or increment in the velocity of a liquid jet accompanied by cavitation in a liquid column (See Figure \ref{type} (c), Supplementary movie 3).
These phenomena are expected to result from interaction of compression and expansion waves with boundaries including a gas-liquid interfaces and the tube walls \citep{Turangan2013}.
The pressure impulse description based on potential flow is therefore inappropriate.

In this paper, we elucidate the mechanisms of the motions of the gas-liquid interfaces induced by a water hammer as displayed in Figure \ref{type}(b)(c). 
For this purpose, we discuss evolution of the pressure waves as well as effects of cavitation that possibly occurs in the liquid column.
We here propose a simple model based on the method of characteristics and compare it with our experiments.


This paper is organized as follows:
In section 2, we show experimental setup and observation.
We propose a model for describing observed phenomena in section 3, followed by comparison with experiments in section 4.
Section 5 concludes our findings.

\section{Experiment}
\label{EM}
\subsection{Experimental setup}
\begin{figure}
\centerline{\includegraphics[scale=0.3]{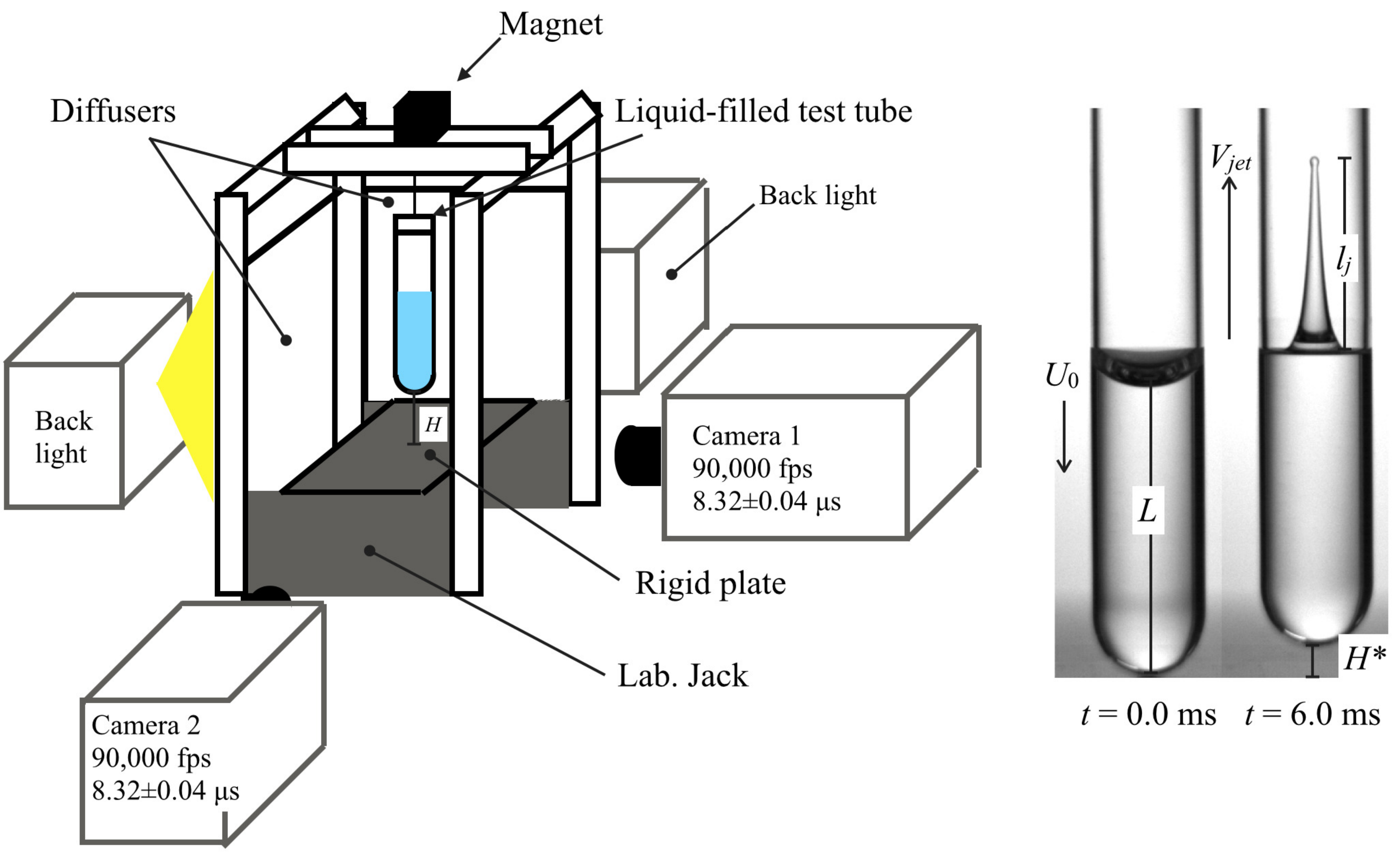}}
  \caption{Schematic of experimental setup}
\label{setup}
\end{figure}

Figure \ref{setup} shows the schematic diagram of our experimental setup in order to observe the motion of the liquid-gas interface by water-hammer events.
We fill a test tube partially with a wetting liquid.
The tube is suspended in a test rig by an electric magnet that touches a dull-shape metal piece on the tube's cap. 
The tube axis aligns with the vertical axis.
When the magnet is turned off, the test tube starts to fall freely from the height $H$ defined as the distance between the tube bottom and the floor.
The gas-liquid interface is in a gravity-free state and deforms as hemispherical shape before the tube impacts a metal plate on a height-adjustable laboratory jack.
The height $H$ is set at a constant value, so that the interface obtains the same shape.
We use two high speed cameras (Photron, Fastcam SA-X) to obtain closeup view of the gas-liquid interface as well as entire view of the liquid column simultaneously.
Frame rate for both cameras is set at 90,000 fps and shutter speed is at 8.32 $\pm$ 0.04 $\mu$s.
Both cameras are triggered by a delay generator (Berkeley Nucleonics, model 575).
The tube is illuminated by white-light sources through diffusers.
All the equipments are placed on a leveled vibration-isolation table (Newport, Smart Table UT2).
We use (gas-saturated) silicone oil (Sigma Aldrich), whose density $\rho_l$, kinematic viscosity $\nu$ and vapor pressure $p_v$ are respectively 930 kg/m$^3$, 10 cSt and 666 Pa at room temperature. 
The test tube is made of borosilicate glass.
The inner diameter and thickness of the tube are 14.2 mm and 1.2 mm, respectively.
The bottom shape of the test tube is rounded, similar to the previous study \citep{Antkowiak2007}.
We summarize experimental parameters in Figure \ref{setup}: the liquid column height $L$, the drop height $H$, and the impact speed $U_0$.
The impact speed $U_0$ can be interpreted as the sum of the drop speed of the tube just before hitting the floor and the rebound speed relative to the floor: $U_0=\sqrt{2gH}+H^\ast/\Delta t_u$.
The drop speed is well approximated by the speed of freely falling bodies $\sqrt{2gH}$ where $g$ denotes the gravitational acceleration (9.81 m/s).
The rebound speed is calculated from the rebound height $H^\ast$ and $\Delta t_u$ (=6.0 ms, see Figure 2).
As summarized in table 1, the impact speed $U_0$ decreases as the liquid column height $L$ increases.
The jet velocity $V_j$ is calculated as $V_j={l_j}/{\Delta t_j}$, where $l_j$ is the jet length and $\Delta t_j = 6.0$ ms (See Figure 2).
We repeat experiments 30 times for each experimental condition.

To classify the phenomena based on whether cavitation occurs in the liquid column, we introduce the cavitation number $Ca$. 
It is defined as $Ca=({P_{atm}-P_v})/{\rho L (a-g)}$, where $P_{atm}$, $P_v$, $\rho$ and $a$ are respectively the atmospheric pressure, the vapor pressure, the liquid density and acceleration imposed on the liquid \citep{Daily2014}.
The cavitation number $Ca$ is the measure of cavitation probability: cavitation is more likely to occur as the value of $Ca$ decreases. 
In our experiments, we judge the occurrence of cavitation if we detect bubbles larger than 1 pixel (=0.16 mm).
The threshold of $Ca < 1$ results in approximately 60 \% of visually detected cavitation bubbles in our experiment in which cavitation nuclei are not controlled but expected to exist randomly on the tube wall and in the liquid column.
The cavitation probability for each experimental condition is also shown in Table \ref{parameter}.

\begin{table}
  \begin{center}
\def~{\hphantom{0}}
  \begin{tabular}{ccccc}
         $L (\mathrm{mm})$  & $H (\mathrm{mm})$ & $U_0 (\mathrm{m/s})$ & $Ca (-)$ & Cavitation probability (\%)\\[3pt]
       30   & 73 & 1.79 $\pm$ 0.02 & 1.02 $\pm$ 0.01 & 0.0 \\
       45   & 73 & 1.78 $\pm$ 0.02 & 0.68 $\pm$ 0.01 & 43.3 \\
       60  & 73 & 1.72 $\pm$ 0.02 & 0.53 $\pm$ 0.01 & 100 \\
       75  & 73 & 1.66 $\pm$ 0.02 & 0.44 $\pm$ 0.01 & 30.0 \\
       90  & 73 & 1.56 $\pm$ 0.02 & 0.39 $\pm$ 0.01 & 60.0 \\
  \end{tabular}
  \caption{Experimental parameters}
  \label{parameter}
  \end{center}
\end{table}
\subsection{Observation}
\label{sec:Observations}
We characterize jet formation based on jet shape and cavitation occurrence inside the liquid column.
To be specific, we categorize all the jets into three types: ``normal-type" jets, ``splash-type" jets, and ``cavitation-type" jets.
In the case of $Ca=1.02$, (see Figure \ref{type}(a) and supplementary movie 1) we obtain a normal type jet as observed in previous experiments \citep{Antkowiak2007, Kiyama2014} in which cavitation was not detected.
On the other hand, for $Ca <$ 1 jet shapes are apparently different from the normal type jet. 
Unless cavitation occurs, there arise non-trivial vibration of the interface and formation of small droplets (Figure \ref{type}(b) and supplementary movie 2); we name this as a splash-type jet.
Once cavitation occurs even in the same value of $Ca$ (but possibly with different state of cavitation nuclei), on the contrary, the jet velocity rises rather beyond jets of the other two types (Figure \ref{type}(c) and supplementary movie 3); we name this as a cavitation-type jet.
Moreover, the spray formation is not obtained in the cavitation type.

\begin{figure}
\centerline{\includegraphics[width=1.0\columnwidth]{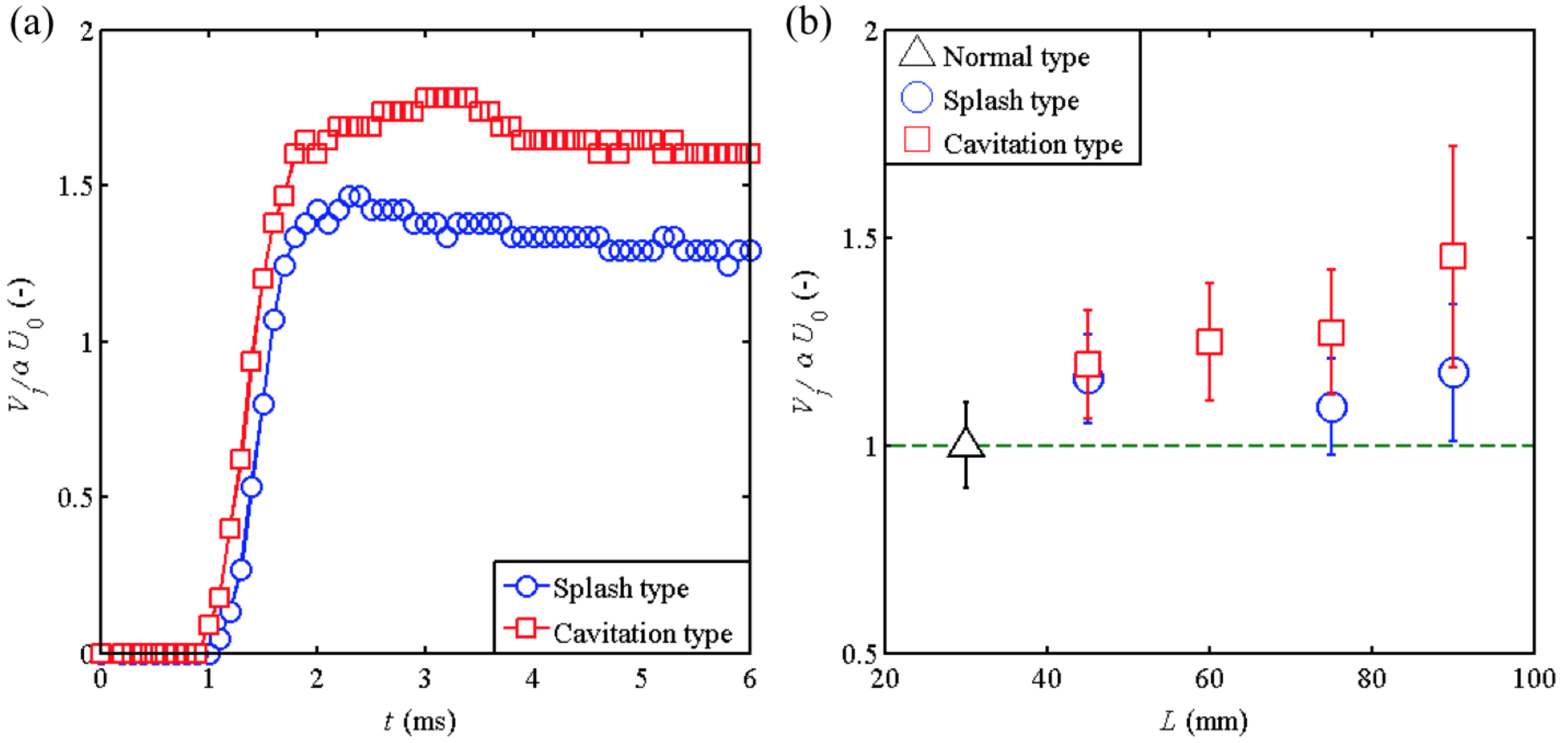}}
  \caption{(a) Temporal evolution of the jet velocities for splash and cavitation types for liquid height $L$=90 mm, (b) The jet velocity vs. the liquid height.}
\label{fig:velocity}
\end{figure}

Figure \ref{fig:velocity}(a) presents temporal evolution of the jet velocities for both splash and cavitation types.
The measured velocity is averaged over $\pm 0.1$ ms to smooth out its fluctuation.
Clearly, there appears a deviation in the jet velocity between the two cases just after the jet formation ($t\approx 1$ ms).
For both cases, the jet velocity reaches its maximum at $t\approx 2$ to $3$ ms and subsequently shows a gradual decay.
We note, for the cavitation type, that cavitation bubbles inside the liquid column collapse at $t\approx 2$ ms as inferred by image analysis.

Figure \ref{fig:velocity}(b) compares the jet velocities for each type as a function of the liquid height $L$.
The vertical axis is the normalized jet velocity $V_j/(\alpha U_{0})$,  
where $\alpha$ is 2.2 so that the jet velocity of the normal-type is scaled as $V_j/(\alpha U_{0})=$ 1.0 ($V_j= 3.74 \pm$0.2 m/s in this particular case).
The jet velocity of the splash type is up to 1.1 times faster than the normal type for all $L$.
In contrast, the jet velocity of the cavitation type for $L=90$ mm is 1.5 times faster than that of the normal type.


\section{Model}
\label{sec:Model}
The unsteady features of the splash- and cavitation-type jets (section 2) cannot be explained by potential flow analysis with the incompressibility constraint.
Thus what we discuss here is the phenomena within a few acoustic time scales (i.e., a period for acoustic wave propagation over the liquid column) $\sim O(0.1)$ ms, which is short compared to the fluid-dynamic time (say, jet formation) $\sim O(1)$ ms.
The aim is to provide the velocity of the gas-liquid interface $U^*$ after a few acoustic time scales during which acoustic waves keep being trapped inside the column through reflections at boundaries.
The flow focusing effect then occurs, leading to the jet velocity $V_j = \alpha U^*$.
For clarity and simplicity, we adopt one-dimensional plane wave model, while ignoring curvature of capillarity as well as the tube bottom with following assumptions: (i) cross-sectional area of the tube is constant throughout the propagation direction; (ii) the tube wall is rigid, (iii) acoustic waves are linear, (iv) the medium is inviscid unless cavitation occurs. 
The extent of fluid-structure interaction may be quantified by the dimensionless parameter $\beta=(c_l^2/c_s^2)(\rho_l/\rho_s)(2R/h)$ where $c$ is the speed of longitudinal sound, $R$ is the mean tube radius, $h$ is the tube thickness, and subscripts $l$ and $s$ denote liquid and solid phases, respectively. In this particular example, we have $\beta\approx 0.1<1$, indicating that tube deformation is expected to be small.
Moreover, pressure perturbations in the liquid up to some hundreds atmospheres remains very weak \citep{Thompson1972}. 
As a result of assumptions (ii) and (iii), acoustic waves in the tube are anticipated to propagate at the speed of sound in the liquid.
Linear wave interactions at boundaries separating different materials can be modeled by the acoustic relation that can be derived from the linearized mass and momentum conservation laws. 
Wave reflections can be quantified based on the acoustic impedance that is a thermodynamic property defined as $I=\rho c$, the product of density $\rho$ and the speed of (longitudinal) sound $c$; the acoustic impedances of gas, liquid and solid phases are denoted by $I_g$, $I_l$ and $I_s$, respectively. 
In the extreme cases where an incident wave comes from very stiff to soft materials (e.g., liquid to gas; $I_l\gg I_g$) and vice versa (e.g., liquid to solid; $I_l\ll I_s$), the acoustic relation becomes very simple. 
If a wave in liquid collides with a gas-liquid interface, the wave transmission to the gas would be so small that pressure at the interface remains almost undisturbed (i.e., free boundary). 
In this case, the interfacial velocity becomes as twice as particle velocity induced by the incident wave. 
On the contrary, even when a wave in liquid collides with a solid boundary, the boundary is essentially fixed (i.e., rigid boundary) and the resulting pressure doubles as a result of superposition of the incident and reflected waves.

\begin{figure}
\centerline{\includegraphics[scale=0.45]{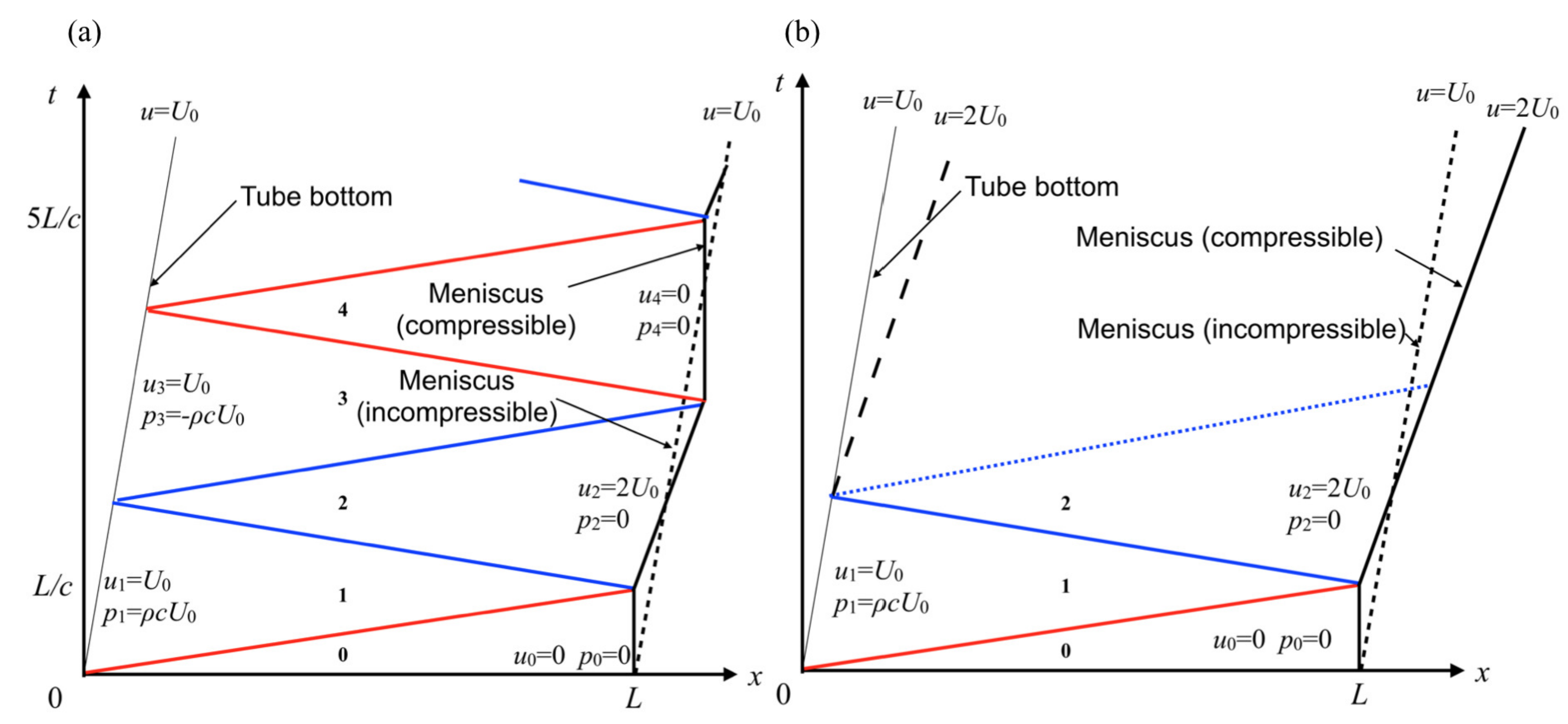}}
  \caption{$x$--$t$ diagrams for (a) Splash-type jet, and (b) Cavitation-type jet. The red and blue solid lines refer a compression and an expansion waves, respectively. The thin black line and the bold black line show position of the tube bottom and the gas-liquid interface, respectively. }
\label{fig:xtdiag}
\end{figure}

With the acoustic relations in these extreme cases, we draw $x$--$t$ diagrams of acoustic wave propagation based on the method of characteristics for jets of normal, splash, and cavitation types in Figure \ref{fig:xtdiag}.
The diagram starts at the moment when the tube wall is set into motion with velocity $U_0$. 
According to assumption (iv), wave attenuation due to dissipative effects is not considered.

 
First we explain how waves evolve in the normal and splash types.
As mentioned in Section 2.2, jet formation for the normal type can be described by pressure impulse in the incompressible sense that the pressure field is built up instantaneously in the entire follow of concern.
Thus, the evolution of the gas-liquid interface starts to move at $U_0$ (see the bold dashed line in Figure \ref{fig:xtdiag} (a)).
For splash-type jets, on the other hand, a pressure wave propagates at the speed of sound in the liquid where cavitation does not occur, and is trapped within the liquid column through multiple reflections; see the red and blue lines in Figure \ref{fig:xtdiag} (a) that denote compression and expansion waves, respectively. 
The induced velocities of the liquid at state 0 to 4 in the diagram are $u_0=0$, $u_1=U_0$, $u_2=2U_0$, $u_3=U_0$ and $u_4=0$, respectively. 
This results in (periodical) vibration of the gas-liquid interface between $u=0$ and $u=2U_0$.
This means that the jet evolves at the average velocity $U_0$ (the same as in the normal jet) but with fluctuation $\pm U_0$ through multiple wave reflections.
The frequency of the interface vibration for the case of $L=90$ mm and $c=990$ m/s is approximated by $c/4L\sim2.8$ kHz.
 
Next, we model cavitation induced by wave interaction in the cavitation-type jet (see Figure \ref{fig:xtdiag} (b)).
Cavitation is expected to occur for liquid pressure to be below a threshold value (e.g., vapor pressure if one ignores surface tension and the dynamics of heterogeneous cavitation nuclei bubbles).
The compression wave initially generated at the tube bottom reaches the gas-liquid interface and reflects as an expansion wave.
The velocity of the gas-liquid interface at this moment is $u=2U_0$.
The expansion wave then reaches the tube bottom and reflects as an expansion wave; unless cavitation occurs, negative pressure in gauge is obtained in the liquid after the expansion wave passes by (i.e., the state 3 in Figure \ref{fig:xtdiag} (a)).
This means that the liquid is stretched and its pressure can possibly be below the cavitation threshold if the initial impact is sufficiently large.
Once cavitation occurs soon after the expansion wave passage, the pressure in the state $3$ in Figure \ref{fig:xtdiag} (a) will be relaxed toward the vapor pressure (or one atmosphere if air dissolved in the liquid comes into cavitation bubbles).
If the expansion wave is significantly damped, the velocity of the gas-liquid interface is expected to be undisturbed at $2U_0$.
Here, the attenuation of the expansion wave may be assumed to be proportional to work done for creation of cavitation bubbles or simply volume of the bubbles.
In this sense, the velocity of the gas-liquid interface $U^*$  could be correlated to cavitation bubble volumes.
We introduce an empirical formula to estimate the velocities of cavitation-type jets as
\begin{equation}
U^*= U_0+C_0\Omega/(S\tau),
\label{eq:CavVj}
\end{equation}
where $C_0$ is a dimensionless fitting constant, $\Omega$ is the maximum volume of cavitation bubbles, $\tau$ is time for bubble growth and $S$ is cross sectional area of the tube.

The maximum volume of cavitation bubbles $\Omega$ is inferred by image analysis.
We treat the bubble as the binarized spot.
We estimate the center of gravity for each bubble, then measure the mean distance in vertical direction, as the typical radius of the bubble, between top/bottom points and the center. 
We take the error as $\pm$ 1 pixel (=0.16 mm) for radii of bubbles.

\section{Comparison}
\label{sec:Comparison}
Based on the wave propagation analysis as well as the empirical formula (Equation \ref{eq:CavVj}) for the cavitation type developed in Section 3, we now reexamine the experimental data.

\begin{figure}
\centerline{\includegraphics[width=1.05\textwidth]{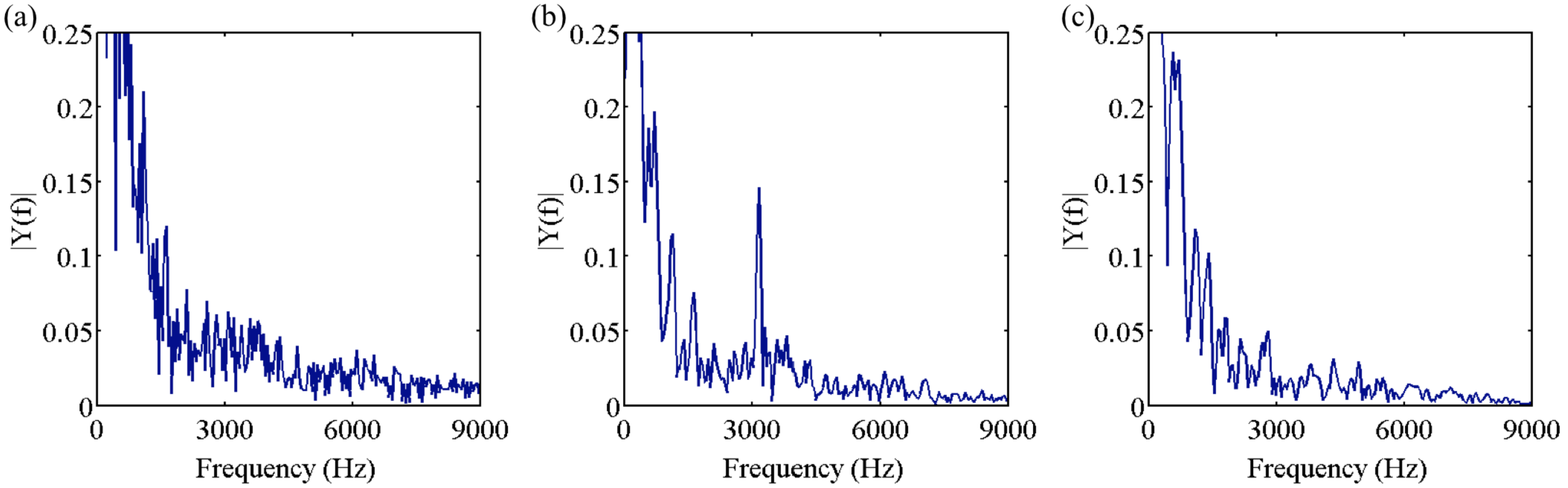}}
  \caption{Frequency analysis of meniscus motion in the neighborhood of the contact line for (a) Normal-type jet, (b) Splash-type jet, and (c) Cavitation-type jet.}
\label{fig:Fig5}
\end{figure}

First, we show frequency analysis of meniscus motion in the neighborhood of the gas-liquid contact line for each type in Figure \ref{fig:Fig5}.
For a splash-type jet in Figure \ref{fig:Fig5}(b), there is a strong peak around 3.1 kHz, which is not observed in other two types. 
This demonstrates multiple reflections of pressure waves trapped between the gas-liquid interface and the tube bottom (See Figure \ref{fig:xtdiag}(a)).
The observed frequency 3.1 kHz show reasonable agreement with the frequency 2.8 kHz predicted in section 3, although the experimental frequency is slightly larger than the model.
The height of the tube's rounded bottom is approximately 9 mm and the radius of curvature of meniscus just before the impact is approximately 7 mm.
Thus the net length of the liquid column for wave propagation may be $\sim$74 mm instead of $L=90$ mm, which results in slightly larger frequency than estimated value 2.8 kHz.
Such a peak does not exist in the cavitation-type jet in Figure 5(c), indicating that the liquid pressure is effectively relaxed to bubble pressure that acoustically hinders wave propagation in the liquid phase.

Second, we compare the empirical formula (Equation \ref{eq:CavVj}) to the measured velocity of cavitation-type jets in Figure 6(a) where the normalized velocity is plotted against the maximum bubble volume to determine the fitting constant in Equation (\ref{eq:CavVj}).
There is a trend, as we expected, that the jet velocity increases as the bubble volume increases.
The fitting constant $C_0$ turns out to be 1.1.
The velocity increment show somewhat linear proportionality with the displaced volume of cavitation bubbles.
It is interesting to note that all the velocity $V_j^\ast/(\alpha U_0)$ is not more than 2.0.
This may also support our model, which predicts the maximum velocity less than 2.0 (see Figure \ref{fig:xtdiag}(b) and Section \ref{sec:Model}). 

\begin{figure}
\centerline{\includegraphics[width=1.05\textwidth]{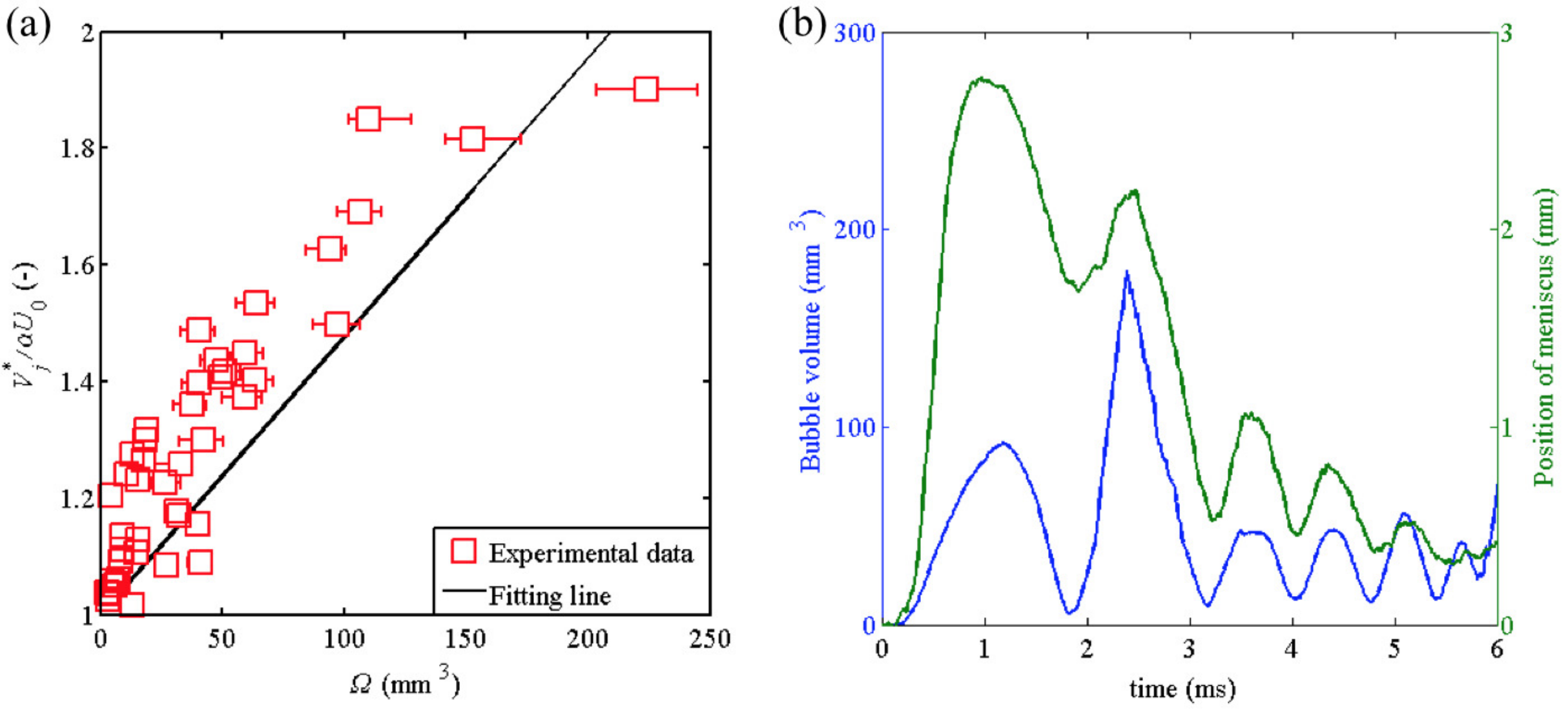}}
  \caption{(a) The velocity of the cavitation-type jet as a  function of the volume of cavitation bubbles, (b) Temporal evolution of the bubble volume and the meniscus position for the cavitation-type.}
\label{fig:Omega_Jet}
\end{figure}

Finally, as displayed in Figure \ref{type}(c), there are crowns near the bottom of the liquid jet.
The position of the bottom of the jet reverses when cavitation bubble collapses (see Figure \ref{fig:Omega_Jet}(b)).
It indicates that the crowns are caused by the secondary shock wave emitted from bubble collapse.



\section{Conclusion}
In this paper, we conducted water-hammer experiments in which a test tube partially filled with a liquid falls under gravity and hits a rigid floor.
We found new types of a jet formed at the gas-liquid interface whose unsteady features cannot be captured by the classical potential flow theory.
We categorized liquid jets into 3 types (normal, splash and cavitation types) based on jet shape and cavitation occurrence inside a liquid column (See Figure \ref{type}).

The splash-type jets showed continuous vibration of the gas-liquid interface while other two types did not.
The velocities of cavitation-type jets were found to be fast compared to these of other two types and accompanied by the onset of cavitation inside the liquid.
In order to understand the phenomena, we proposed a new model for explaining pressure wave propagation and its interaction with the boundaries and the effect of cavitation bubbles.
For splash-type jets, the vibration of the interface was caused by repeated wave reflection within the liquid column. 
Thus the vibration frequency can be estimated by the liquid column height and the speed of sound.
For cavitation-type jets, we considered the attenuation of expansion waves due to pressure relaxation around cavitation bubbles, which leads to the emergence of a faster liquid jet.
We speculated that the jet velocity of the cavitation-type can be correlated to the displaced volume of bubbles (Equation \ref{eq:CavVj}).
We compared the vibration frequency of the gas-liquid interface for splash-type jet and found a reasonable agreement between experiments and the model (see Figure \ref{fig:Fig5}). 
We also compared the empirical formula (Equation \ref{eq:CavVj}) with the experiments and found, as we expected, that the velocity increment for the cavitation-type jet can be well estimated by the displaced volume of cavitation bubbles (Figure \ref{fig:Omega_Jet}).
In short, the proposed model that accounts for acoustics and cavitation can properly explain the new types of focused liquid jets.

We thank Y. Watanabe, M. Maeshima and K. Hirose for helping our experiments. 
This work was supported by JSPS KAKENHI Grant Number 26709007. 

\bibliographystyle{jfm}

\bibliography{jfm-instructions}

\end{document}